\documentclass[english,prd,twocolumn,superscriptaddress,nofootinbib,preprintnumbers]{revtex4}
\usepackage[latin1]{inputenc}
\usepackage{graphicx}
\usepackage{amssymb}

\makeatletter

\usepackage{amsfonts}
\usepackage{dcolumn}
\usepackage{hyperref}

\def\be{\begin{equation}}
\def\ee{\end{equation}}
\def\ba{\begin{eqnarray}}
\def\ea{\end{eqnarray}}
\def\bs{\begin{subequations}}
\def\es{\end{subequations}}

\newcommand{\rd}{{\rm d}}

\usepackage{color}

\makeatother

\usepackage{babel}
\makeatother
\begin{document}

\title{Observational signatures of $f(R)$ dark energy models \\
that satisfy cosmological and local gravity constraints}

\author{Shinji Tsujikawa}

\affiliation{Department of Physics, Gunma National College of Technology, Gunma
371-8530, Japan}

\begin{abstract}

We discuss observational consequences of $f(R)$ dark energy scenarios
that satisfy local gravity constraints (LGC) as well as conditions of the 
cosmological viability. The model we study is given by  
$m(r)=C(-r-1)^p$ ($C>0, p>1$) with $m=Rf_{,RR}/f_{,R}$ 
and $r=-Rf_{,R}/f$, which cover viable $f(R)$ models 
proposed so far in a high-curvature region designed to 
be compatible with LGC. The equation of state of dark energy 
exhibits a divergence at a redshift $z_c$ that can be as close 
as a few while satisfying sound horizon 
constraints of Cosmic Microwave Background (CMB).
We study the evolution of matter density perturbations 
in details and place constraints on model parameters from the
difference of spectral indices of power spectra 
between CMB and galaxy clustering.
The models with $p \ge 5$ can be consistent with 
those observational constraints as well as LGC.
We also discuss the evolution of perturbations in the 
Ricci scalar $R$ and show that an oscillating mode
(scalaron) can easily dominate over a matter-induced mode
as we go back to the past. 
This violates the stability of cosmological 
solutions, thus posing a problem about how the over-production
of scalarons should be avoided in the early universe. 

\end{abstract}

\date{\today}

\maketitle

%%%%%%%%%%%%%%
\section{Introduction}
%%%%%%%%%%%%%%

The origin of dark energy (DE) has persistently been one of the 
most serious problems in cosmology \cite{review,CST}. 
Many DE models have been proposed so far, 
but we have not found any strong evidence
to support that such models are better than 
cosmological constant.
Thus the first step towards the understanding of the origin of DE 
is to find the departure from the $\Lambda$CDM model.

The simplest modification to the $\Lambda$CDM model is 
perhaps so-called $f(R)$ gravity in which the Lagrangian 
is written in terms of the function of a Ricci scalar $R$.
It is well known that inflationary expansion is realized
by the Starobinsky's model with a Lagrangian density 
$f(R)=R+\alpha R^2$ \cite{star}.
Since the $R^2$ term is negligibly small relative to $R$
at the present epoch, this model is not suitable to explain 
present accelerated expansion of the Universe.
Instead the model with a Lagrangian density 
$f(R)=R-\alpha/R^n$ ($\alpha>0, n>0$)
was proposed to give rise to a late-time accelerated 
expansion in the metric formalism \cite{fR} 
(see also Refs.~\cite{Capoluca,early}).
However it was shown that this model is plagued by 
a matter instability \cite{Dolgov}
as well as by a difficulty to satisfy local gravity 
constraints \cite{LG}.
Moreover it does not possess a standard matter-dominated epoch 
because of a large coupling between dark energy and 
dark matter \cite{APT} (see Refs.~\cite{recent} for recent works).

In Ref.~\cite{AGPT} several conditions for the cosmological viability 
of $f(R)$ dark energy models were derived without specifying 
the forms of $f(R)$. This can be well understood by considering 
a trajectory of each model in the $(r,m)$ plane, where 
$r \equiv -Rf_{,R}/f$ and $m \equiv Rf_{,RR}/f_{,R}$.
The existence of a saddle matter-dominated epoch 
requires the conditions $m>0$ and $-1<{\rm d}m/{\rm d}r \le 0$
around the point $(r,m)=(-1,0)$.
The matter era can be followed by a stable de-Sitter 
attractor on the line $r=-2$ provided that $0<m(r=-2) \le 1$.
This method is useful to rule out some of $f(R)$
models such as $f(R)=R-\alpha/R^n$ 
($\alpha>0, n>0$) easily.

More recently a sequence of cosmologically viable $f(R)$ 
models was discussed in Ref.~\cite{Li,AT07}.
One of such models, for example, is 
$f(R)=(R^b-\Lambda)^{1/b}$ with $0<b<1$, 
which corresponds to a straight line 
$m(r)=(b-1)(r+1)$ connecting the matter point 
$(r,m)=(-1,0)$ to the de-Sitter point on the line 
$r=-2$. The parameter $m$ that characterizes the deviation 
from the $\Lambda$CDM model is constrained to be 
$m<{\cal O}(0.1)$ for such models 
from the data of Supernova Ia (SN Ia) 
and Cosmic Microwave Background (CMB) \cite{AT07}.
Meanwhile local gravity experiments constrain the value of $m$
to be very much smaller than unity in high-density regions 
where gravity experiments are carried out.
This means that the deviation from the $\Lambda$CDM
model needs to be very small in a high-curvature cosmological 
epoch whose Ricci scalar $R$ is much larger than 
the present cosmological value $R_0$.

A number of authors \cite{Hu,star07,Appleby} recently proposed $f(R)$
dark energy models that can satisfy both cosmological 
and local gravity constraints (LGC) by using the so-called 
chameleon mechanism \cite{KW}
(see Refs.~\cite{Olmo,lgcpapers} for related works).
{}From the requirement of LGC these behave as close as 
the $\Lambda$CDM model during radiation and matter 
dominated epochs ($R \gg R_0$).
The deviation from the $\Lambda$CDM model becomes
significant after the end of the matter era with the growth 
of the quantity $m$. These models satisfy the relation 
$f(R=0)=0$, implying that cosmological constant disappears
in a flat spacetime. We note, however, that the Ricci scalar
is frozen at a value $R=R_1>0$ if the solutions are trapped 
by stable de-Sitter attractors responsible for the late-time 
acceleration. Thus in these models the system does not 
reach the region $R=0$ in an asymptotic future.

In this paper we shall study observational consequences of 
$f(R)$ models that satisfy LGC in addition to conditions 
of cosmological viability. The models we consider are given by 
$m(r)=C(-r-1)^p$ with $C>0$ and $p>1$, which cover 
viable models proposed in literature \cite{Hu,star07,Appleby}
in the region $R \gg R_0$.
The quantity $m(r)$ is, for large $p$, vanishingly small
during matter and radiation epochs ($r \approx-1$), 
but grows to the order of $C$ as the solutions approach 
de-Sitter attractors on the line $r=-2$.
These models exhibit peculiar evolution of the 
DE equation of state, as we will see later.
Moreover matter density perturbations evolve differently 
compared to the $\Lambda$CDM cosmology for redshifts 
below a critical value $z_k$. This property can be used to place 
constraints on model parameters in addition to constraints
coming from LGC, SN Ia and CMB.

This paper is organized as follows.
In Sec.~\ref{moex} we present all conditions viable 
$f(R)$ DE models need to satisfy.
In addition to $f(R)$ models studied so far, we shall propose 
another model satisfying these conditions.
In Sec.~\ref{SNI} the evolution of DE equation of state and 
resulting observational consequences are discussed in addition to 
constraints coming from the sound horizon of CMB.
In Sec.~\ref{matter} we study how matter 
perturbations evolve on sub-horizon scales and put constraints
on model parameters from the difference of spectral indices of 
the power spectra between CMB and galaxy clustering.
We also discuss the evolution of the perturbation $\delta R$
and show that an oscillating mode called scalaron \cite{star} 
easily dominates over the background value $R$ 
when we go back to the past.
This generally violates the stability condition of $f(R)$ models, 
which gives rise to another problem about how to avoid the 
over-production of scalarons in the early universe.
We conclude in Sec.~\ref{conclude}.

%%%%%%%%%%%%%%%%%%%%%%%%%%%%%%%%%%%%%%%
\section{Models that satisfy cosmological and local gravity constraints}
\label{moex}
%%%%%%%%%%%%%%%%%%%%%%%%%%%%%%%%%%%%%%%

Let us begin with the following action
\begin{equation}
\label{action}
S=\int{\rm d}^{4}x\sqrt{-g}\left[\frac{1}{2\kappa^{2}}f(R)
+{\mathcal{L}}_{{\rm m}}+{\mathcal{L}}_{{\rm rad}}\right]\,,
\end{equation}
where $\kappa^{2}=8\pi G$ ($G$ is a bare gravitational constant). 
In what follows we use the unit $\kappa^2=1$, but we restore the 
gravitational constant when it is needed.
Note that ${\mathcal{L}}_{{\rm m}}$ and ${\mathcal{L}}_{{\rm rad}}$ 
are the Lagrangian densities of dust-like
matter and radiation, respectively, which satisfy 
usual conservation equations.
In the flat Friedmann-Robertson-Walker (FRW) background 
with a scale factor $a$ 
the Ricci scalar is given by $R=6(2H^2+\dot{H})$,
where $H \equiv \dot{a}/a$ is a Hubble parameter 
and a dot represents a derivative with respect to cosmic 
time $t$.

There are a number of constraints viable $f(R)$ models need to 
satisfy. First of all, to avoid anti-gravity, we require the condition 
$f_{,R} \equiv {\rm d}f/{\rm d}R>0$.
Modified $f(R)$ models possess a scalar particle 
whose effective mass is given by 
\begin{eqnarray}
\label{Mdef}
M^2(R) \simeq \frac{1}{3f_{,RR}}\,, 
\end{eqnarray}
in the regime $M^2(R) \gg R$ \cite{AT07,star07,Fara}.
In order to avoid that the scalaron becomes tachyons or ghosts, 
we require $f_{,RR} \equiv {\rm d}^2f/{\rm d}R^2>0$ in this region.
Note that this condition can be also derived by considering 
the stability of perturbations \cite{Fara,Song}.

The conditions for the cosmological viability of $f(R)$ models 
have been studied in 
Ref.~\cite{AGPT} in great details. 
This can be well understood by considering two quantities:
\begin{eqnarray}
\label{mdef}
m = \frac{Rf_{,RR}}{f_{,R}}\,,~~~~~
r= -\frac{Rf_{,R}}{f}\,.
\label{ldef}
\end{eqnarray}
The $\Lambda$CDM model, $f(R)=R-2\Lambda$, corresponds to
$m=0$ and $r=-R/(R-2\Lambda)$.
The quantity $m$ characterizes the deviation from the 
$\Lambda$CDM model. 
The cosmological viability of such models is known by 
plotting corresponding curves in the $(r,m)$ plane. 

In what follows we shall consider cosmological evolution that starts from 
a radiation epoch with large and positive $R$ followed by a matter era and 
eventually approaches a de-Sitter attractor with $R=R_1>0$ 
in future\footnote{During the radiation era the Ricci scalar evolves as 
$R \propto t^{-3/2}$ because of the presence of non-relativistic particles.}.
In cosmologically viable models we study, the quantity $m$ is always 
smaller than 1 with $f_{,R}$ of order unity before reaching a de-Sitter attractor.
Since $1/f_{,RR} \gg R$ in such cases, one can use the scalaron mass 
given in Eq.~(\ref{Mdef}). Hence the stability conditions are given by \cite{star07} 
\begin{eqnarray}
\label{con1}
f_{,R}>0\,,~~~f_{,RR}>0\,,~~{\rm for}~~R \ge R_1\,.
\end{eqnarray}

The matter-dominated point, $P_M$, exists on the line 
$m=-r-1$ with $m$ close to 0, i.e., $(r,m) \approx (-1, 0)$.
The presence of a viable saddle matter era
demands the conditions \cite{AGPT}  
(see also Ref.~\cite{rbean}):
\begin{eqnarray}
\label{con3}
m(r \approx -1)>0\,,~~~{\rm and}~~~
-1<\frac{{\rm d}m}{{\rm d}r}(r \approx -1) \le 0\,.
\end{eqnarray}
If the condition (\ref{con1}) is satisfied then the variable 
$m$ is automatically positive.
The second requirement in Eq.~(\ref{con3}) implies that 
$m(r)$ curves should be be present between the lines 
$m=0$ and $m=-r-1$.

There is a stable de-Sitter fixed point that leads to
a late-time acceleration:
\begin{eqnarray}
\label{con4}
P_{A} :~r=-2\,,~~0 <m \le 1\,.
\end{eqnarray}
If a $m(r)$ curve staring from $P_M$ has an intersection point 
with a line $r=-2$ in the region $0 <m \le 1$, the corresponding 
$f(R)$ model is regarded as cosmologically viable.
The $\Lambda$CDM model is a straight line that links $P_M$: $(r,m)=(-1,0)$ 
with $P_{A}$: $(r,m)=(-2,0)$.
In this paper we do not consider another accelerated fixed point 
$P_B$ that exists on the line $m=-r-1$ with 
$(\sqrt{3}-1)/2<m \le 1$ \cite{AGPT}. 
This corresponds to the case in 
which $R$ continues to decrease in future, which can 
violate the stability condition (\ref{con1}).

A number of $f(R)$ models satisfying the above conditions were considered 
in Refs.~\cite{Li,AT07}. Some examples are 
\begin{eqnarray}
\label{fRluca1}
&&{\rm (i)}~f(R)=(R^{b}-\Lambda)^{c}~~~(c\ge1,~~bc \approx 1)\,,\\
\label{fRluca2}
&&{\rm (ii)}~f(R)=R-\alpha R^{n}~~~(\alpha>0,~~0<n<1)\,,
\end{eqnarray}
which correspond to $m(r)=[(1-c)/c]r+b-1$ and 
$m(r)=n(1+r)/r$, respectively. 

Let us next consider local gravity constraints 
on $f(R)$ dark energy models.
The LGC are satisfied for $M \ell \gg 1$ \cite{Olmo,AT07}, 
where $\ell$ is a scale at which gravity experiments are carried out.
Using Eqs.~(\ref{Mdef}) and (\ref{mdef}), this constraint is expressed by 
\begin{eqnarray}
m(R_s) \ll \frac{1}{f_{,R_s}} \left( \frac{\ell}{R_s^{-1/2}} \right)^2\,,
\end{eqnarray}
where $R_s$ is a curvature measured on the local structure and is 
proportional to the energy density $\rho_s$ of the structure 
($R_s \approx 8\pi G \rho_s$). 
Using the present cosmological density $\rho_0$ and the Hubble radius
$H_0^{-1} \sim 10^{28}$ cm (in what follows we use the 
subscript ``0'' for present values), the above constraint is rewritten as
\begin{eqnarray}
\label{lgccon}
m(R_s) \ll \frac{\rho_s}{\rho_0}
\left( \frac{\ell}{H_0^{-1}} \right)^2\,,
\end{eqnarray}
where we used $f_{,R_s} \sim 1$ and $R_0 \sim 
H_0^2 \sim 8\pi G \rho_0$.
The r.h.s. of Eq.~(\ref{lgccon}) is very much smaller 
than unity \cite{AT07} because $\ell \ll H_0^{-1}$ 
even though $\rho_s$ is larger than $\rho_0$.
In the case of the Cavendish-type experiments
the typical constraint is $m(R_s) \ll 10^{-43}$, 
as we will see later.

The above argument shows that in the high-curvature region ($R \gg R_0$)
the quantity $m$ needs to be negligibly small.
Cosmologically this means that during radiation and matter eras
the models need to mimic the $\Lambda$CDM model with a high-precision.
Note that the models (i) and (ii) given in Eqs.~(\ref{fRluca1}) and 
(\ref{fRluca2}) behave as 
$m(r)=C(-r-1)$ as $r$ approaches $-1$. 
In such cases, however, LGC are not satisfied unless $C$ is chosen 
to be unnaturally small. 

Hu and Sawicki \cite{Hu} proposed an explicit $f(R)$ model 
that satisfies both cosmological and local gravity constraints.
It is given by 
\begin{eqnarray}
\label{mo1}
f(R)=R-\lambda R_c \frac{(R/R_c)^{2n}}{(R/R_c)^{2n}+1}\,,
\end{eqnarray}
where the power $2n$ is used instead of $n$.
Starobinsky \cite{star07} also proposed another viable model:
\begin{eqnarray}
\label{mo2}
f(R)=R-\lambda R_c \left[ 1-\left( 1+\frac{R^2}{R_c^2}
\right)^{-n} \right]\,.
\end{eqnarray}
In both models $n$, $\lambda$ and $R_c$ are positive constants, 
where $R_c$ is the order of the present Ricci scalar $R_0$.
Since $f(R=0)=0$ cosmological constant 
disappears in a flat spacetime.
Thus the origin of dark energy can be regarded as the geometrical one.

Let us check the cosmological viability as well as the stability 
for such models.  
In the region $R \gg R_c$ these behave as
\begin{eqnarray}
\hspace*{-0.2em}f(R) &\simeq& R-\lambda R_c \left[1-
\left( \frac{R_c}{R}  \right)^{2n} \right]\,, \\ 
\label{rra}
\hspace*{-0.2em}r &\simeq& -1-\lambda \frac{R_c}{R}\,,\\
\hspace*{-0.2em} m &\simeq& \frac{2n(2n+1)}{\lambda^{2n}}
(-r-1)^{2n+1}\,. 
\end{eqnarray}
Thus in this region the models (\ref{mo1}) and (\ref{mo2}) 
have the following property
\begin{eqnarray}
\label{ourmodel}
m(r)=C(-r-1)^p\,,
\end{eqnarray}
where $p=2n+1>1$ and $C$ is a positive constant.
It is obvious that,  for larger $p$, $m(r)$ becomes very small as $r \to -1$ 
so that the model satisfies LGC.
Since $\frac{{\rm d}m}{{\rm d}r}(r=-1)=0$ the condition (\ref{con3}) is also satisfied.
 
Let us next check the conditions (\ref{con1}) and (\ref{con4}). 
In the model (\ref{mo1}), the de-Sitter point at $r=-2$ is determined 
by the value of $\lambda$:
\begin{eqnarray}
\lambda=\frac{(1+x_1^{2n})^2}{x_1^{2n-1}
(2+2x_1^{2n}-2n)}\,,
\end{eqnarray}
where $x_1=R_1/R_c$. {}From the stability condition $0<m(r=-2) \le 1$
we obtain
\begin{eqnarray}
\label{hucon}
\hspace*{-0.4em}2x_1^{4n}-(2n-1)(2n+4)x_1^{2n}
+(2n-1)(2n-2) \ge 0\,.
\end{eqnarray}
When $n=1$, for example, we have 
$x_1 \ge \sqrt{3}$ and $\lambda \ge 8\sqrt{3}/9$.
Under Eq.~(\ref{hucon}) one can show that the condition 
(\ref{con1}) is satisfied.
This situation is similar to the Starobinsky's model (\ref{mo2}), see 
Ref.~\cite{star07} for details.

We can extend the above two models to the more general form
\begin{eqnarray}
\hspace*{-0.2em}f(R)=R-\xi(R),~~\xi(0)=0,
~~\xi(R \gg R_c) \to {\rm const}.
\end{eqnarray}
The conditions (\ref{con1}) translate into   
\begin{eqnarray}
\xi_{,R}<1\,,~~\xi_{,RR}<0\,,~~{\rm for}~~R \ge R_1\,.
\end{eqnarray}
In order to satisfy LGC, we require that $\xi(R)$
approaches a constant rapidly as $R$ grows in the region
$R \gg R_c$ (such as $\xi(R) \simeq {\rm constant}-(R_c/R)^{2n}$ 
discussed above). Another model to meet these requirements is 
\begin{eqnarray}
\label{mo3}
f(R)=R-\lambda R_c {\rm tanh}\,\left( \frac{R}{R_c}\right)\,,
\end{eqnarray}
where $\lambda$ and $R_c$ are positive constants.
A similar model was proposed by Appleby and Battye \cite{Appleby}, 
although it is different from (\ref{mo3}) in the sense that $\xi(R)$ can be 
negative for $R<R_1$.
In the region $R \gg R_c$ the model (\ref{mo3}) behaves as 
$f(R) \simeq R-\lambda R_c[1-\exp(-2R/R_c)]$, which can be regarded 
as the special case of (\ref{ourmodel}) with a limit $p \to \infty$.
The Ricci scalar at the de-Sitter point is determined by $\lambda$, as
\begin{eqnarray}
\lambda=\frac{x_1\,{\rm cos h}^2(x_1)}
{2\,{\rm sin h}(x_1)\,{\rm cos h}(x_1)-x_1}\,,
\end{eqnarray}
where $x_1=R_1/R_c$. {}From the stability 
condition (\ref{con4}) of the de-Sitter point 
we obtain the constraint 
\begin{eqnarray}
\label{cb}
x_1 > 0.920\,,~~~\lambda > 0.905\,.
\end{eqnarray}

\begin{figure}
\begin{centering}\includegraphics[width=3.4in,height=3.2in]{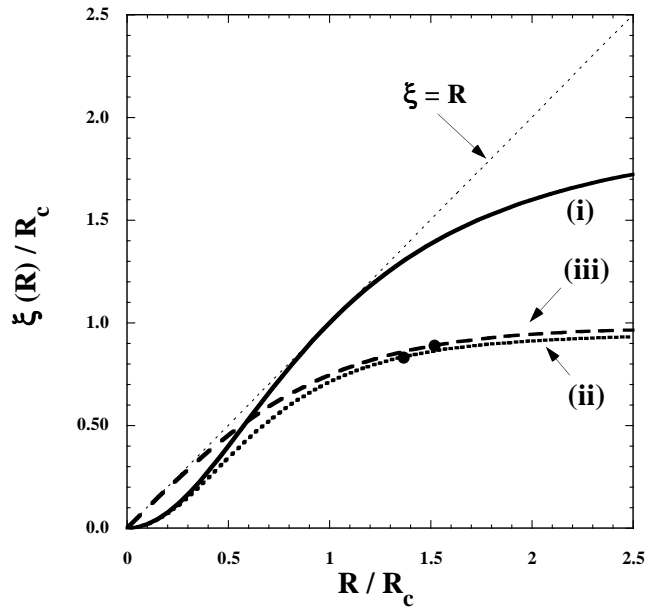} \par\end{centering}
\caption{\label{fig1} The illustration of $\xi(R)$ as a function of $R/R_c$ 
in three different models. Each corresponds to (i) the model (\ref{mo1}) by Hu and Sawicki
with $n=1$ and $\lambda=2$, (ii) the model (\ref{mo2}) by Starobinsky with $n=2$ 
and $\lambda=0.95$, and (iii) the model (\ref{mo3}) with $\lambda=0.98$.
We also plot a line $\xi(R)=R$ to see whether or not the condition $\xi_{,R}<1$ 
(i.e., $f_{,R}>0$) is violated. 
The black points represent de-Sitter fixed points ($R=R_1$). Note that 
in the case (i) the de-Sitter point corresponds to $R_1/R_c=3.383$, which 
is outside of the figure.
In the case (i) we have $f_{,R}<0$ for 
$0.296<R/R_c<1$, whereas in the cases (ii) and (iii) 
$f_{,R}>0$ for all positive $R$.
In the models (\ref{mo1}) and (\ref{mo2}) it is inevitable to avoid that $f_{,RR}$ 
becomes negative in the small $R$ region, 
but in the model (\ref{mo3}) it is possible to realize $f_{,RR}>0$
for $R>0$. In the region before the solutions reach the de-Sitter 
attractor (i.e., $R \ge R_1$), both $f_{,R}$ and $f_{,RR}$ are positive 
in the above three models.
}
\end{figure}

In the models (\ref{mo1}) and (\ref{mo2}), $f_{,RR}$ are negative 
for $0<R/R_c<[(2n-1)/(2n+1)]^{1/2n}$ and $0<R/R_c<1/\sqrt{2n+1}$,
respectively. In the model  (\ref{mo1}) the quantity $f_{,R}$ also becomes 
negative (i.e., $\xi_{,R}>1$) in some regions, whereas in the model (\ref{mo2})
it is possible to have $f_{,R}>0$ for all positive $R$ in 
some restricted regions of the parameter space ($0.944<\lambda<0.966$ for $n=2$).
See the curves (i) and (ii) in Fig.~\ref{fig1} for illustration.
Of course we are considering the situation in which 
the violation of the conditions $f_{,R}>0$ and $f_{,RR}>0$ occurs for $R<R_1$, 
so it is harmless as long as the universe is in the region $R \ge R_1$.  
In the model (\ref{mo3}) we always have $f_{,RR}>0$ for positive $R$, whereas
$f_{,R}$ is positive for $\lambda<1$.
Hence, if $0.905<\lambda<1$, this model satisfies the conditions
$f_{,R}>0$ and $f_{,RR}>0$ for all positive $R$ and also 
possesses a de-Sitter attractor.
Such an example is plotted as the case (iii) in Fig.~\ref{fig1}. 

\begin{figure}
\begin{centering}\includegraphics[width=3.4in,height=3.2in]{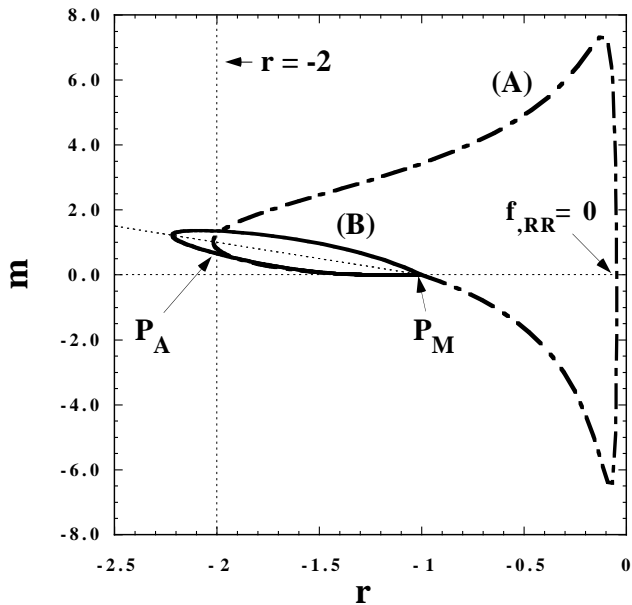} \par\end{centering}
\caption{\label{fig2} Two trajectories in the $(r,m)$ plane.
The trajectory (A) corresponds to the model (\ref{mo2}) by Starobinsky 
with $n=2$ and $\lambda=0.95$, whereas the trajectory (B) to 
the model (\ref{mo3}) $\lambda=0.95$.
In both cases the solutions start from the region around the point 
$P_M: (r,m)=(-1,0)$ with $R \gg R_c$. 
They approach the stable de-Sitter point 
$P_A$ on the line $r=-2$ with $0<m \le 1$. 
In the case (A) the quantity $f_{,RR}$
becomes negative for $0<R/R_c<1/\sqrt{2n+1}$, 
while $f_{,R}$ is positive for $R>0$.
In the case (B) one has $f_{,RR}>0$ and $f_{,R}>0$
for all positive $R$. In the limit $R/R_c \to 0$ both models approach 
the point $(r,m)=(-1,0)$ again.
}
\end{figure}

In Fig.~\ref{fig2} we plot the trajectories in the $(r,m)$ plane
for the model (\ref{mo2}) with $n=2$, $\lambda=0.95$, and 
for the model (\ref{mo3}) with $\lambda=0.95$.
The solutions start from the region $R/R_c \gg 1$ around the point $P_M$ 
and they finally approach the de-Sitter point $P_A$ at $R=R_1$.
Since $f_{,R}>0$ ($R>0$) and $f(0)=0$ in such cases 
we have $f(R)>0$ for positive $R$, 
which means that $r=-Rf_{,R}/f$ is always negative. 
The quantity $m=Rf_{,RR}/f_{,R}$ is positive for $R \ge R_1$
because $f_{,RR}>0$.
In the case (A) of Fig.~\ref{fig2} we have $m<0$ for 
$0<R/R_c<1/\sqrt{2n+1}$ because $f_{,RR}$ changes the sign.
Meanwhile, in the case (B), $m$ is always positive for $R>0$.   
In the limit $R/R_c \to 0$ the trajectories 
approach the point $(r,m)=(-1,0)$ again in both models. 
If the Ricci scalar oscillates around $R=0$,
the quantity $f_{,RR}$ becomes negative for $R<0$ even for the model (\ref{mo3}).
As we will see later in details, this can indeed occur by the oscillation 
of scalarons unless initial conditions are appropriately chosen.
The above argument shows the importance of confining the Ricci scalar 
in the region $R \ge R_1$ to ensure the stability of models.

In this paper we shall study a number of cosmological constraints 
on the models of the type 
(\ref{ourmodel}). As we mentioned, in the region $R \gg R_c$, 
this covers the models (\ref{mo1}) and (\ref{mo2}) with $p=2n+1$
as well as the model (\ref{mo3}) with the limit $p \to \infty$.
If $0<C \le 1$ there exists a stable de-Sitter point $P_A$ at $r=-2$.
The model $m(r)=C(-r-1)^p$ essentially contains sufficient information 
about how viable $f(R)$ models behave. During radiation and matter 
eras ($r \simeq -1$) the quantity $m$ is very much smaller than unity, 
but it grows to the order of $C$ once the system approaches the
de-Sitter attractor $P_A$.
Thus one can see the departure from the 
$\Lambda$CDM model around the present epoch.  
Note that we are only concerned with the region of $m(r)$ curves
before the solutions reach the de-Sitter attractor.
If we demand the condition $f(R=0)=0$, this can be satisfied
by modifying the form of $m(r)$ outside the region that connects
$P_M$ to $P_A$ (as in the models (\ref{mo2}) and (\ref{mo3})
 in Fig.~\ref{fig2}).

Before entering the details of various cosmological constraints, 
we consider LGC on the model (\ref{ourmodel}).
In high-dense regions where local gravity experiments
are carried out ($R_s \gg R_c \sim R_0$) the quantity $r$ behaves as 
Eq.~(\ref{rra}) and hence $m \simeq \tilde{m} (R_0/R_s)^p$, where 
$\tilde{m}$ is a constant whose order is not much different from unity.
Since $R_0/R_s \sim \rho_0/\rho_s$, the 
constraint (\ref{lgccon}) yields
\begin{eqnarray}
\label{lgc2}
\left( \frac{\rho_s}{\rho_0} \right)^{p+1}
\gg \tilde{m} \left(\frac{H_0^{-1}}{\ell} \right)^2\,.
\end{eqnarray}
In the Cavendish-type experiments the typical values are
$\rho_s \sim 10^{-12}$ g/cm$^3$ and 
$\ell \sim 10^{-2}$ cm \cite{hoyle}.
Recalling the values $\rho_0 \sim 10^{-29}$ g/cm$^3$
and $H_0^{-1} \sim 10^{28}$ cm, we find that the LGC 
is well satisfied for $p \ge 3$.

In the case of solar-system experiments, if we take the typical solar-system 
length scale $\ell=1$\,Au$=1.5 \times 10^{13}$ cm with the density 
$\rho_s \sim 10^{-24}$ g/cm$^3$ at the distance from Sun \cite{Hu}, 
we obtain the constraint $p \ge 5$.
Meanwhile the so-called Shapiro time-delay effect \cite{Shapiro}
comes mainly from the gravity contribution around the radius of 
Sun ($\ell \sim 7.0 \times 10^{10}$ cm) with the density 
$\rho_s \sim 10^{-15}$ g/cm$^3$, which gives a much weaker
constraint: $p \ge 2$.
Thus the constraint $p \ge 5$ is certainly enough
to satisfy LGC and is even too tight 
in some of gravity experiments. 

%%%%%%%%%%%%%%%%%%%%%%%%%%%%
\section{SN Ia and CMB sound horizon constraints}
\label{SNI}
%%%%%%%%%%%%%%%%%%%%%%%%%%%%

In this section we discuss the cosmological evolution
of the model (\ref{ourmodel}) at the background level and
confront it with constraints coming from SN Ia and 
the sound horizon of CMB.
In the flat FRW spacetime the variation of the action (\ref{action})
leads to the following equations
\begin{eqnarray}
3FH^{2} & = & \rho_{{\rm m}}+\rho_{{\rm rad}}
+(FR-f)/2-3H\dot{F},\label{E1}\\
-2F\dot{H} & = & \rho_{{\rm m}}+(4/3)\rho_{{\rm rad}}
+\ddot{F}-H\dot{F},\label{E2}
\end{eqnarray}
where $F \equiv \partial f/\partial R$.
Here $\rho_{\rm m}$ and $\rho_{\rm rad}$ are 
the energy densities of a non-relativistic matter and radiation, 
respectively, which satisfy the usual conservation equations.

Following Refs.~\cite{AGPT,AT07} we introduce 
the dimensionless variables
\begin{eqnarray}
& &x_{1}=-\frac{\dot{F}}{HF}\,,~~x_{2}=-\frac{f}{6FH^{2}}\,,
\nonumber \\
& &x_{3}=\frac{R}{6H^{2}}\,,~~x_{4}
=\frac{\rho_{{\rm rad}}}{3FH^{2}}\,.
\label{didef}
\end{eqnarray}
Then we obtain the dynamical equations \cite{AGPT}
\begin{eqnarray}
x_{1}' & = & -1-x_{3}-3x_{2}+x_{1}^{2}-x_{1}x_{3}+x_{4}~,\label{N1}\\
x_{2}' & = & \frac{x_{1}x_{3}}{m}-x_{2}(2x_{3}-4-x_{1})~,\label{N2}\\
x_{3}' & = & -\frac{x_{1}x_{3}}{m}-2x_{3}(x_{3}-2)~,\label{N3}\\
x_{4}' & = & -2x_{3}x_{4}+x_{1}\, x_{4}\,,\label{N4}
\end{eqnarray}
where a prime represents a derivative with respect to $N=\ln a$.
Since $m$ is a function of $r=x_3/x_2$, the above system is closed.
The energy density and the pressure of DE to confront with SN Ia observations
are given in Ref.~\cite{AT07} and the corresponding 
equation of state (EOS) of DE is 
\begin{eqnarray}
\label{wde}
w_{\rm DE}=-\frac{1}{3}\frac{2x_{3}-1
+(F/F_0)\,x_{4}}{1-(F/F_0)(1-x_{1}-x_{2}-x_{3}-x_4)}\,,
\end{eqnarray}
where $F_0$ is the present value.

Our model (\ref{ourmodel}) needs to satisfy the condition $F_{,R}>0$
for $R \ge R_1$, which leads to the increase of $F$ toward the past as 
$R$ gets larger. The denominator in Eq.~(\ref{wde}) is written as 
$1-(F/F_0)\Omega_{\rm m}$, where 
\begin{eqnarray}
\Omega_{\rm m} \equiv \frac{\rho_m}{3FH^2}
=1-x_1-x_2-x_3-x_4\,.
\end{eqnarray}
Since $\Omega_{\rm m}$ increases 
from present to the matter-dominated epoch, 
it happens that $w_{\rm DE}$ exhibits a divergence at a redshift 
$z_c$ satisfying $\Omega_{\rm m}=F_0/F$. 
This is in fact generic to cosmologically viable models 
that fulfill the criterion (\ref{con1}). 
The EOS of DE crosses a cosmological constant  
boundary ($w_{\rm DE}=-1$) at a redshift $z_b$ 
smaller than $z_c$ \cite{AT07}.

\begin{figure}
\begin{centering}\includegraphics[width=3.4in,height=3.2in]{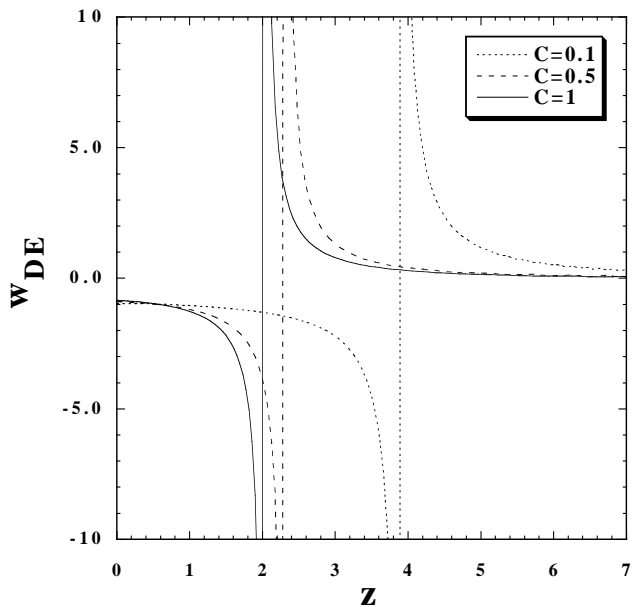} \par\end{centering}
\caption{\label{fig3} The DE equation of state $w_{{\rm DE}}$ versus the 
redshift $z$ for the model $m(r)=C(-r-1)^2$ with three different 
parameters ($C=0.1,0.5,1.0$).
With the increase of $C$ the divergence 
of $w_{{\rm DE}}$ occurs for smaller $z$.}
\end{figure}

In Fig.~\ref{fig3} we plot $w_{\rm DE}$ versus the redshift 
$z  \equiv a_0/a-1$ for the model $m(r)=C(-r-1)^2$
with three different values of $C$. 
In all simulations the present epoch ($z=0$) is identified as
a matter energy fraction $\Omega_{\rm m}=0.28$ with 
a radiation contribution $\Omega_{\rm rad} 
\equiv \rho_{\rm rad}/3FH^2 \sim 10^{-4}$. 
Note that the current universe is in the middle
of approaching the de-Sitter attractor 
$P_A$: $(x_1, x_2, x_3)=(0, -1, 2)$ from 
the matter point $P_M$: $(x_1, x_2, x_3) \approx (0, -1/2, 1/2)$.
As we see in Fig.~\ref{fig3}, $w_{\rm DE}$ is larger than $-1$ at $z=0$ and decreases
to $-\infty$ as $z \to z_c-0$ after crossing the cosmological constant boundary
at $z=z_b$.
For smaller $C$, $z_c$ gets larger. When $C=1$ we have $z_c=2.03$
and $w_{\rm DE}(z=0)=-0.853$, whereas if $C=0.5$ we get 
$z_c=2.31$ with $w_{\rm DE}(z=0)=-0.904$.
This peculiar behavior of the EOS of DE is an interesting signature 
to discriminate $f(R)$ models from the $\Lambda$CDM cosmology.
In particular the larger deviation from the $\Lambda$CDM model leads to 
a smaller critical redshift $z_c$ that can be reached in future observations.

For the values of $p$ greater than 2, $z_c$ gets larger and $w_{\rm DE}(z=0)$
tends to be closer to $-1$. In Table I we show $z_b$, $z_c$ and 
$w_{\rm DE}(z=0)$ together with present values of $m$ for 
several different choices of $p$ and $C$.
When $p \ge 5$, if we look at the low-redshift region only, 
the models are hardly distinguishable from 
the $\Lambda$CDM model.
Still the EOS of DE shows a peculiar behavior in the high-redshift region: 
$z>3$. For larger $p$ the cosmological constant 
boundary crossing occurs at the redshift close to the present epoch.
We note that the recent SN Ia data analysis finds some evidence
for such a crossing \cite{Nese}.
If future high-precision observations favour models whose EOS 
corresponds to a phantom ($w_{\rm DE}<-1$) 
in most of the past epochs relevant to SN Ia observations,
this can be the signal of $f(R)$ gravity.

If we use the criterion $w_{{\rm DE}}(z=0)<-0.7$ according to the 
current SN Ia data \cite{Astier}, we find from Table I that the models with 
$p \ge 2$ satisfy this requirement even when $C$ is as close as unity.
The slopes of the EOS, $|{\rm d}w_{{\rm DE}}/{\rm d}z|$,
are found to be smaller than the order of 0.1 around the present epoch,
which do not provide additional information to constrain models.
When $C=1$, we have $w_{{\rm DE}}(z=0)>-0.7$ only for $p \le 1.4$.
Thus the current SN Ia observations do not provide a better constraint 
on the power $p$ than the one obtained by LGC. 
It will be interesting, however, to carry out a likelihood analysis
using the future data of SN Ia along the line of Refs.~\cite{Arman}.

%%%%%%%%%%%%%%%%%%%%%%%%%%%%%
\begin{table}[t]
\begin{tabular}{|c|c|c|c|c|c|}
\hline 
$p$ &
$C$ &
$z_{b}$ &
$z_{c}$ &
$w_{{\rm DE}}(z=0)$ &
$m(z=0)$\tabularnewline
\hline 
$1.5$ &
$0.1$ &
$0.79$ &
$3.44$ &
$-0.959$ &
$0.075$ \tabularnewline
\hline 
$1.5$ &
$0.5$ &
$0.73$ &
$2.24$ &
$-0.851$ &
$0.244$ \tabularnewline
\hline 
$1.5$ &
$1$ &
$0.85$ &
$2.14$ &
$-0.743$ &
$0.293$ \tabularnewline
\hline 
$2$ &
$0.1$ &
$0.62$ &
$3.88$ &
$-0.969$ &
$0.067$ \tabularnewline
\hline 
$2$ &
$0.5$ &
$0.54$ &
$2.31$ &
$-0.904$ &
$0.220$\tabularnewline
\hline 
$2$ &
$1$ &
$0.56$ &
$2.03$ &
$-0.853$ &
$0.285$\tabularnewline
\hline 
$3$ &
$0.1$ &
$0.40$ &
$5.00$ &
$-0.982$ &
$0.055$\tabularnewline
\hline 
$3$ &
$0.5$ &
$0.33$ &
$2.83$ &
$-0.955$ &
$0.180$\tabularnewline
\hline
$3$ &
$1$ &
$0.32$ &
$2.31$ &
$-0.936$ &
$0.248$\tabularnewline
\hline
$5$ &
$0.1$ &
$0.16$ &
$7.25$ &
$-0.994$ &
$0.037$\tabularnewline
\hline
$5$ &
$0.5$ &
$0.10$ &
$4.30$ &
$-0.990$ &
$0.124$\tabularnewline
\hline
$5$ &
$1$ &
$0.09$ &
$3.53$ &
$-0.988$ &
$0.180$\tabularnewline
\hline
\end{tabular}
\caption[table1]{\label{table1}  The values $z_{b}$, $z_{c}$, $w_{{\rm DE}}(z=0)$
and $m(z=0)$ in the model $m(r)=C(-r-1)^p$ for several different choices of 
$p$ and $C$.}
\end{table}
%%%%%%%%%%%%%%%%%%%%%%%%

Let us also consider the sound horizon constraint coming from CMB.
The angular size of the sound horizon is defined by 
\begin{eqnarray}
\Theta_{s}=\int_{z_{{\rm {dec}}}}^{\infty}\frac{c_{s}(z){\rm {d}}z}
{H(z)}\,\Biggl/\int_{0}^{z_{{\rm {dec}}}}\frac{{\rm {d}}z}{H(z)}\,,
\end{eqnarray}
where $c_{s}^{2}(z)=1/[3(1+3\rho_{b}/4\rho_{\gamma})]$ is the adiabatic
baryon-photon sound speed and $z_{{\rm {dec}}}\simeq1089$.
{}From the position of CMB acoustic peaks we obtain the constraint
$\Theta_{s}=0.5946\pm0.0021$ deg from the WMAP 3-year data \cite{Spergel}.
For the models in which the effect of dark energy is not negligible 
during the matter-dominated epoch, the quantity $\Theta_{s}$ is rather 
strongly modified (as in the coupled quintessence \cite{coupled}).

In $f(R)$ gravity, if the quantity $m$ is not much smaller than 1
during the matter era, this leads to a considerable change 
of $\Theta_{s}$ compared to the $\Lambda$CDM model. 
In fact this happens for the models (\ref{fRluca1}) 
and (\ref{fRluca2}), as was shown in Ref.~\cite{AT07}. 
In our model (\ref{ourmodel}) the quantity $m$ is 
very much smaller than unity during the matter era from the requirement 
to satisfy LGC. Hence it is easier to satisfy the sound horizon constraint
compared to the models (\ref{fRluca1}) and (\ref{fRluca2}).
In fact we have evaluated $\Theta_{s}$ numerically and confirmed that 
the models with $p \ge 2$ are consistent with the  WMAP 3-year data.
Thus the data coming from the CMB sound horizon does not provide 
tighter constraints relative to the SN Ia data.

%%%%%%%%%%%%%%%%%%%%%%%%%%%%
\section{Matter perturbations and scalaron oscillations}
\label{matter}
%%%%%%%%%%%%%%%%%%%%%%%%%%%%

In this section we study constraints on the model (\ref{ourmodel}) 
coming from matter density perturbations.
Let us consider scalar metric perturbations $\alpha$, 
$\beta$, $\varphi$ and $\gamma$ about the flat 
FRW background \cite{metric}:
\begin{eqnarray}
\rd s^2 &=&-(1+2\alpha) \rd t^2-2a \beta_{,i}
\rd t \rd x^{i} \nonumber \\
& &+a^2 \left[(1+2\varphi) 
\delta_{ij}+2\gamma_{|ij} \right] \rd x^i \rd x^j\,.
\end{eqnarray}
In what follows we neglect the contribution of radiation as it is 
unimportant to discuss the evolution of matter perturbations 
during the matter-dominated epoch.
The energy-momentum tensors of a pressureless matter is 
decomposed by 
\begin{eqnarray}
T^{0}_{0}=-(\rho_m+\delta \rho_m)\,,\quad
T^0_{i}=-\rho_mv_{m,i}\,,
\end{eqnarray}
where $v_m$ is a velocity potential.

Introducing a covariant velocity perturbation, $v \equiv av_m$, we obtain 
the following equations of motion in the Fourier space \cite{Hwang}
(see also Refs.~\cite{Song,linear}):
\begin{widetext}
\begin{eqnarray}
\label{pereq1}
& & \alpha=\dot{v}\,, \\
\label{pereq2}
& & (\delta \rho_m/\rho_m)^{\cdot}=\kappa-3H\alpha-
\frac{k^2}{a^2}v\,, \\
\label{pereq3}
& & \dot{\kappa}+2H\kappa+\left( 3\dot{H}-
\frac{k^2}{a^2} \right) \alpha
=\frac{1}{2F} \left[ \left(-6H^2+\frac{k^2}{a^2}
\right) \delta F+3H\delta \dot{F} +3\delta \ddot{F}
-\dot{F}\kappa-3(2\ddot{F}+H\dot{F})\alpha
-3\dot{F}\dot{\alpha}+\delta \rho_m \right], \\
\label{pereq4}
& & \delta \ddot{F}+3H\delta \dot{F}+\left(\frac{k^2}
{a^2} -\frac{R}{3} \right)\delta F
=\frac13 \delta \rho_m+\dot{F} (\kappa+\dot{\alpha})
+(2\ddot{F}+3H\dot{F})\alpha-\frac13 F \delta R,\\
\label{pereq5}
& & -\frac{k^2}{a^2}\varphi+3H(H\alpha-\dot{\varphi})
+\frac{k^2}{a^2}H \chi=\frac{1}{2F} \left[
3H \delta \dot{F}-\left(3\dot{H}+3H^2-\frac{k^2}{a^2}
\right) \delta F-3H \dot{F} \alpha -\dot{F} \kappa
-\delta \rho_m \right]\,,\\
\label{pereq6}
& & \dot{\chi}+H \chi-\alpha-\varphi=
\frac{1}{F} ( \delta F-\dot{F}\chi)\,,
\end{eqnarray}
\end{widetext}
where $k$ is a comoving wavenumber and 
$\kappa \equiv 3(H\alpha-\dot{\varphi})+
(\beta+a\dot{\gamma})k^2/a$.
We define a gauge-invariant quantity: 
$\delta_m \equiv \delta \rho_m/\rho_m+3Hv$.
In the comoving gauge where $v=0$, we find from 
Eqs.~(\ref{pereq1}) and (\ref{pereq2}) that 
$\alpha=0$ and $\kappa=\dot{\delta}_m$.
Then from Eqs.~(\ref{pereq3}) and (\ref{pereq4}) 
we obtain 
\begin{widetext}
\begin{eqnarray}
\label{delm}
& & \ddot{\delta}_m+\left(2H+\frac{\dot{F}}{2F}
\right) \dot{\delta}_m-\frac{\rho_m}{2F}\delta_m
=\frac{1}{2F} \left[ \left(-6H^2+\frac{k^2}{a^2}
\right)\delta F+3H\delta \dot{F}+3\delta \ddot{F}
\right]\,,\\
\label{delF}
& & \delta \ddot{F}+3H\delta \dot{F}+\left(\frac{k^2}
{a^2}+\frac{F}{3F_{,R}}-4H^2-2\dot{H} \right)\delta F
=\frac13 \delta \rho_m+\dot{F} \dot{\delta}_m\,.
\end{eqnarray}
\end{widetext}

In the model (\ref{ourmodel}) the quantity $m=Rf_{,RR}/F$ is very much 
smaller than unity during the  matter era with $F \simeq 1$. 
Since $1/f_{,RR} \gg R$ in such a case, the scalaron mass squared 
is given by Eq.~(\ref{Mdef}) and satisfies the relation 
$M^2 \gg R \sim H^2$. In what follows we shall discuss
two cases: (A) $M^2 \gg k^2/a^2$ and (B) $M^2 \ll k^2/a^2$, 
separately. As we will see below, the modes that are 
initially in the region (A) can enter the region (B) during the 
matter-dominated epoch.

\subsection{The region $M^2 \gg k^2/a^2$}

When $M^2 \gg k^2/a^2$, Eq.~(\ref{delF}) is approximately 
given by 
\begin{eqnarray}
\label{delFap}
\delta \ddot{F}+3H\delta \dot{F}+M^2\delta F
 \simeq \frac13 \delta \rho_m\,,
\end{eqnarray}
where we used the fact that the variation of the quantity $F$
is negligibly small during the matter era.
This is a very good approximation for the model (\ref{ourmodel}),
since $m$ is vanishingly small during the matter era.

The general solutions for Eq.~(\ref{delFap}) are given by 
the sum of the oscillating solution $\delta F_{\rm osc}$ 
obtained by setting $\delta \rho_m=0$ and the special 
solution $\delta F_{\rm ind}$ of Eq.~(\ref{delFap})
induced by the presence of matter 
perturbations $\delta \rho_m$.
The former was obtained by Starobinsky \cite{star07}
for the model (\ref{mo2}) in the unperturbed 
flat FRW background (i.e., $k=0$).
The oscillating part $\delta F_{\rm osc}$ satisfies 
the equation $(a^{3/2} \delta F_{\rm osc})^{\cdot \cdot}+
M^2(a^{3/2} \delta F_{\rm osc}) \simeq 0$. 
By using the WKB approximation, we obtain the solution  
\begin{eqnarray}
\delta F_{\rm osc} \propto
a^{-3/2}\,f_{,RR}{}^{1/4}\,\cos 
\left( \int \frac{1}{\sqrt{3f_{,RR}}} {\rm d}t  
\right)\,.
\end{eqnarray}

During the matter era in which the background Ricci scalar 
evolves as $R^{(0)}=4/(3t^2)$, 
the quantity $f_{,RR}$ has a dependence 
$f_{,RR} \propto R^{-(p+1)} \propto t^{2(p+1)}$.
Hence the evolution of the perturbation, $\delta R_{\rm osc}=
\delta F_{\rm osc}/f_{,RR}$, is given by 
\begin{eqnarray}
\label{delos}
\delta R_{\rm osc} \simeq c \,t^{-\frac{3p+5}{2}}\, 
\cos( c_0 \,t^{-p})\,,
\end{eqnarray}
where $c$ and $c_0$ are constants.
As we go back to the past the amplitude of $\delta R_{\rm osc}$
dominates over $R^{(0)}$, unless the coefficient $c$ is chosen to 
be very small. 
Since $R$ gets smaller than $R_1$ and even becomes 
negative, the stability condition (\ref{con1}) 
is violated.
This property also holds in the radiation era
during which $\delta R_{\rm osc}$ and the background 
Ricci scalar $R^{(0)}$ evolve as
\begin{eqnarray}
\delta R_{\rm osc} \simeq c\,t^{-\frac{9p+15}{8}} 
\cos(c_0 \,t^{-\frac14 (3p-1)})\,,~~~
R^{(0)} \propto t^{-3/2}\,.
\end{eqnarray}
Thus we need to avoid the excessive production of scalarons in the 
early universe so that $|\delta R_{\rm osc}| \ll R^{(0)}$ is satisfied
at all times. This problem is even severe for the models 
of the type (\ref{mo3}). 
Moreover the scalaron mass rapidly grows to the past 
in these models and can exceed the Planck mass even 
during the matter era.

\begin{figure}
\begin{centering}\includegraphics[width=3.4in,height=3.2in]{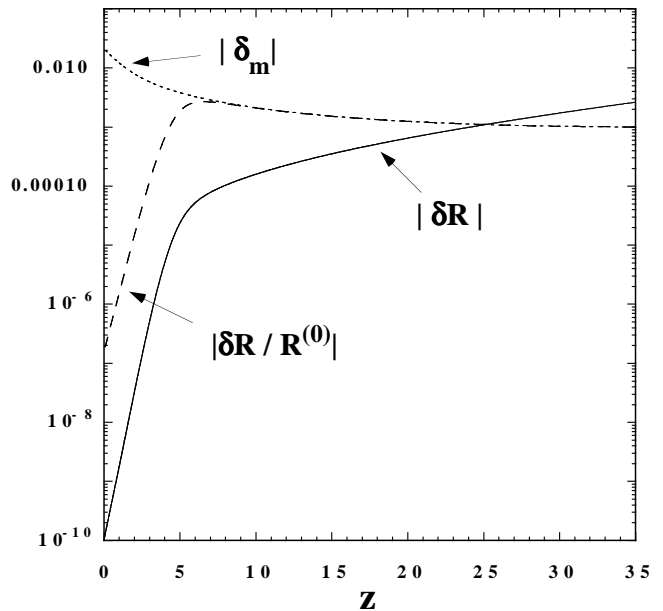} 
\par\end{centering}
\caption{\label{fig4} The evolution of $\delta R$, $\delta R/R^{(0)}$ 
and $\delta_m$ for the model $m(r)=(-r-1)^3$ with the mode 
$k/a_0H_0=335$ in the case where the coefficient $b_1$ 
in Eq.~(\ref{Rratio}) is very small
so that the scalaron mode $\delta R_{\rm osc}$ is negligible relative to the matter 
induced mode $\delta R_{\rm ind}$.
The transition from the region $M^2 \gg k^2/a^2$ to the region 
$M^2 \ll k^2/a^2$ occurs around the redshift $z_k=5$.
}
\end{figure}

The special solution $\delta F_{\rm ind}$ of Eq.~(\ref{delFap}) 
can be derived by using the approximation used in 
Refs.~\cite{efp,Tsujiper}. This amounts to neglecting the 
first and second terms relative to others, giving 
\begin{eqnarray}
\label{Find}
\delta F_{\rm ind} \simeq f_{,RR}\,\delta \rho_m\,,\quad
\delta R_{\rm ind} \simeq \delta \rho_m\,.
\end{eqnarray}
Under the condition $|\delta F_{\rm osc}| \ll |\delta F_{\rm ind}|$ we 
have $\delta F \simeq f_{,RR}\,\delta \rho_m$.
Substituting this relation for Eq.~(\ref{delm}) and 
using the property $M^2 \gg k^2/a^2$, we obtain 
\begin{eqnarray}
\label{delsta}
\ddot{\delta}_m+2H \dot{\delta}_m
-4\pi G \rho_m \delta_m \simeq 0\,.
\end{eqnarray}
Here we have reproduced the gravitational constant for clarity.
This is the usual equation of matter perturbations on sub-horizon modes
in $\Lambda$CDM cosmology and has a growing 
mode solution $\delta_m \propto a \propto t^{2/3}$.
{}From Eq.~(\ref{Find}) we get 
\begin{eqnarray}
\delta F_{\rm ind} \propto t^{2p+2/3}\,,\quad
\delta R_{\rm ind} \propto t^{-4/3}\,.
\end{eqnarray}
Compared to the oscillating mode (\ref{delos}), the induced matter
mode $\delta R_{\rm ind}$ decreases more slowly and thus 
dominates in the late universe.
Relative to the background value $R^{(0)}$, 
the perturbation, $\delta R=\delta R_{\rm osc}+\delta R_{\rm ind}$, 
evolves as
\begin{eqnarray}
\label{Rratio}
\frac{\delta R}{R^{(0)}} \simeq b_1\,t^{-\frac{3p+1}{2}}
\cos (c_0 t^{-p})+b_2\,t^{2/3}\,,
\end{eqnarray}
where $b_1$ and $b_2$ are constants. 
Unless the coefficient $b_1$ is very small, 
the oscillating mode dominates over the matter induced mode 
to violate the condition $R \ge R_1$ as we go back to the past.
In Fig.~\ref{fig4} we show an example about the evolution of perturbations 
in which the initial condition of $\delta R$ is chosen to be very close to 
$\delta \rho_m$ (see Appendix for perturbation equations
suitable for numerical calculations). 
The perturbation evolves as $\delta R \propto t^{-4/3}$
during the period in which the condition $M^2 \gg k^2/a^2$
is satisfied. After the system enters the region $M^2 \ll k^2/a^2$,
$\delta R$ decreases more rapidly as we  
see in the next subsection.
Since $\delta R \simeq  \delta \rho_m$ and $R^{(0)} 
\simeq 3H^2 \simeq \rho_m$ during the matter era,  
we obtain the relation $\delta R/R^{(0)} \simeq \delta_m$. 
This property is in fact confirmed in Fig.~\ref{fig4} in the 
region $M^2 \gg k^2/a^2$.

\begin{figure}
\begin{centering}\includegraphics[width=3.4in,height=3.2in]{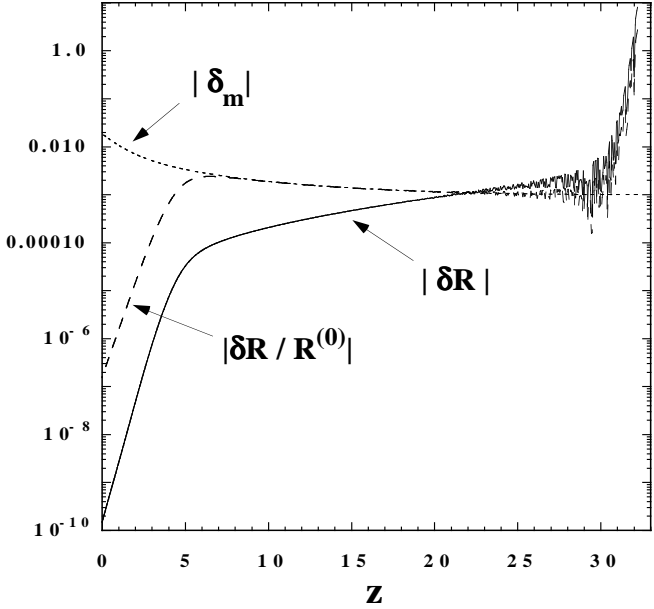} 
\par\end{centering}
\caption{\label{fig5} The evolution of $\delta R$, $\delta R/R^{(0)}$ 
and $\delta_m$ for the model $m(r)=(-r-1)^3$ with the mode 
$k/a_0H_0=315$ in the case where the coefficient $b_1$ 
in Eq.~(\ref{Rratio}) is not chosen to be very small.
The scalaron mode $\delta R_{\rm osc}$ 
dominates over the matter induced mode 
$\delta R_{\rm ind}$ around the redshift $z \gtrsim 30$.}
\end{figure}

Figure \ref{fig5} is the case in which the oscillating mode
dominates over $\delta R_{\rm ind}$ around the redshift $z \gtrsim 30$.
Since $|\delta R|$ grows to the order of $R^{(0)}$ the Ricci scalar $R$
becomes negative in this region, thus violating the stability condition
(\ref{con1}). These results confirm that the coefficient $b_1$ 
should be chosen to be very small to avoid the dominance of the 
scalaron mode in an early epoch.

\subsection{The region $M^2 \ll k^2/a^2$}

Since the scalaron mass decreases as $M \propto t^{-(p+1)}$, 
the modes which initially exist in the region $M^2 \gg k^2/a^2$ 
can enter the regime $M^2 \ll k^2/a^2$ during 
the matter-dominated epoch.
In this regime Eq.\,(\ref{delF}) is approximately given by 
\begin{eqnarray}
\label{delFap2}
\delta \ddot{F}+3H\delta \dot{F}+\frac{k^2}{a^2}\delta F
 \simeq \frac13 \delta \rho_m.
\end{eqnarray}
Using the WKB approximation, the solution corresponding to 
the scalaron mode is
\begin{eqnarray}
\delta R_{\rm osc}=\frac{\delta F_{\rm osc}}{f_{,RR}}
 \simeq c\,t^{-2p-8/3} \cos ( c_0 k t^{1/3})\,.
\end{eqnarray}
The matter-induced special solution of Eq.~(\ref{delFap2}) 
is approximately given by
\begin{eqnarray}
\delta F_{\rm ind} \simeq \frac{a^2}{3k^2}\delta \rho_m\,.
\end{eqnarray}

{}From Eq.~(\ref{delm}) we obtain the following approximate equation  
under the condition $|\delta F_{\rm osc}| \ll |\delta F_{\rm ind}|$:
\begin{eqnarray}
\label{delm2}
\ddot{\delta}_m+2H \dot{\delta}_m
-\frac43 \cdot 4\pi G \rho_m \delta_m \simeq 0\,.
\end{eqnarray}
Relative to the region $M^2 \gg k^2/a^2$
the growth rate of $\delta_m$ is enhanced and is given by 
\begin{eqnarray}
\label{delmap}
\delta_m \propto t^{\frac{\sqrt{33}-1}{6}}\,.
\end{eqnarray}
Hence the induced-matter mode evolves as 
\begin{eqnarray}
\delta F_{\rm ind} \propto t^{\frac{\sqrt{33}-5}{6}}\,,\quad
\delta R_{\rm ind} \propto t^{-2p+\frac{\sqrt{33}-17}{6}}\,.
\end{eqnarray}
Then the evolution of the perturbation 
$\delta R=\delta R_{\rm osc}+\delta R_{\rm ind}$, 
relative to $R^{(0)}$, is given by 
\begin{eqnarray}
\label{delR2}
\frac{\delta R}{R^{(0)}} \simeq b_1 t^{-2p-2/3} 
\cos(c_0 k t^{1/3})+b_2 t^{-2p+\frac{\sqrt{33}-5}{6}}\,.
\end{eqnarray}
As long as the scalaron mode is suppressed at the beginning 
of the matter era, the second term on the r.h.s. of Eq.~(\ref{delR2})
dominates over the first one.
In Figs.~\ref{fig4} and \ref{fig5} the sudden decrease of 
$\delta R$ means that the system enters the region 
$M^2 \ll k^2/a^2$ in which the evolution of $\delta R$
is characterized by 
$\delta R \propto t^{-2p+\frac{\sqrt{33}-17}{6}}$.
At this stage $\delta R/R^{(0)}$ is no longer proportional 
to $\delta_m$.

\subsection{The matter power spectra}

The evolution of the matter perturbation is 
given by $\delta_m \propto t^{2/3}$ for $M^2 \gg k^2/a^2$
and $\delta_m \propto t^{(\sqrt{33}-1)/6}$ 
for $M^2 \ll k^2/a^2$. 
We shall use the subscript ``$k$'' for the quantities at which
$k$ is equal to $aM$, whereas the subscript ``$\Lambda$''
is used at which the accelerated expansion starts ($\ddot{a}=0$).
While the redshift $z_\Lambda$ is independent of $k$, 
$z_k$ depend on $k$ and also on the mass $M$.

%%%%%%%%%%%%%%%%%%%%%%%%%%%%%
\begin{table}[t]
\begin{tabular}{|c|c|c|c|c|c|}
\hline 
$p$ &
$z_\Lambda$ &
$z_k$ &
$\delta n^{({\rm A})} (t_\Lambda)$ &
$\delta n^{({\rm N})} (t_\Lambda)$ &
$\delta n^{({\rm N})} (t_0)$\tabularnewline
\hline 
$2$ &
$0.95$ &
$9.62$ &
$0.106$ &
$0.107$ &
$0.108$ \tabularnewline
\hline 
$3$ &
$0.86$ &
$4.83$ &
$0.075$ &
$0.074$ &
$0.077$\tabularnewline
\hline 
$4$ &
$0.81$ &
$3.25$ &
$0.057$ &
$0.056$ &
$0.061$\tabularnewline
\hline 
$5$ &
$0.78$ &
$2.49$ &
$0.047$ &
$0.045$ &
$0.053$\tabularnewline
\hline 
$6$ &
$0.76$ &
$2.03$ &
$0.039$ &
$0.035$ &
$0.044$\tabularnewline
\hline
$7$ &
$0.75$ &
$1.72$ &
$0.034$ &
$0.028$ &
$0.039$\tabularnewline
\hline
\end{tabular}
\caption[table2]{\label{table2} The redshifts $z_{\Lambda}$ 
and $z_{k}$ in the model $m(r)=(-r-1)^p$ for the mode
$k/a_0H_0=300$.
We also show analytic and numerical values of $\delta n (t_\Lambda)$,
which are denoted as $\delta n^{({\rm A})} (t_\Lambda)$ and 
$\delta n^{({\rm N})}  (t_\Lambda)$ respectively.
The values $\delta n^{({\rm N})}  (t_0)$ are
obtained by numerically integrating perturbation equations 
up to the present epoch.
}
\end{table}
%%%%%%%%%%%%%%%%%%%%%%%%

%
\begin{figure}
\begin{centering}\includegraphics[width=3.4in,height=3.2in]{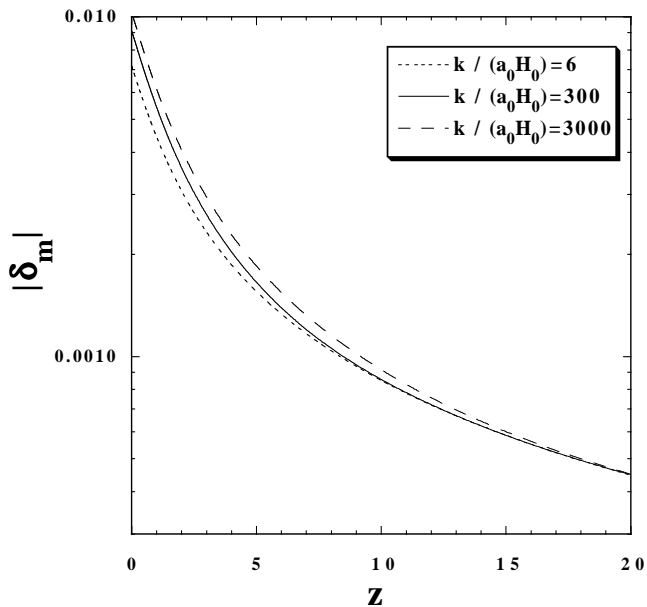} 
\par\end{centering}
\caption{\label{fig6} The evolution of the matter perturbation $\delta_m$
in the model $m(r)=(-r-1)^2$ for the modes $k/a_0H_0=6, 300, 3000$.
The transition redshift $z_k$ increases for larger $k$. For the mode
$k/a_0H_0=300$ we have $z_k=9.62$.}
\end{figure}

In Table II we show numerical values of $z_\Lambda$ and 
$z_k$ for the mode $k/a_0H_0=300$ in the model 
$m(r)=(-r-1)^p$. 
For smaller $p$ the period of 
a non-standard evolution of 
$\delta_m$ becomes longer because $z_k$
tends to be larger.
We also note that, for larger $k$, $z_k$ gets larger.
This means that the duration of the
period of an additional amplification of $\delta_m$
is different depending on the mode $k$, 
see Fig.~\ref{fig6}.
Since the time $t_k$ has a dependence 
$t_k \propto k^{-\frac{3}{3p+1}}$, the matter power spectrum 
$P_{\delta_m}=(k^3/2\pi^2)|\delta_m|^2$ 
at the time $t_\Lambda$ shows a difference compared to the case of 
the $\Lambda$CDM model:
\begin{eqnarray}
\frac{P_{\delta_m}(t_\Lambda)}
{P_{\delta_m}{}^{\Lambda{\rm CDM}}(t_\Lambda)}
=\left(\frac{t_\Lambda}{t_k}\right)
^{2\left(\frac{\sqrt{33}-1}{6}-\frac23\right)}
\propto k^{\frac{\sqrt{33}-5}{3p+1}}\,.
\end{eqnarray}

While the galaxy matter power spectrum is modified by 
this effect, the CMB spectrum is hardly affected
except for low multipoles around which an integrated 
Sachs-Wolfe (ISW) effect becomes important \cite{SPH}.
Thus there is a difference for the spectral indices
of two power spectra, i.e., 
\begin{eqnarray}
\label{deln}
\delta n (t_\Lambda)=\frac{\sqrt{33}-5}{3p+1}\,.
\end{eqnarray}

Since $z_k$ becomes as close as $z_\Lambda$
for larger $p$, it is not necessarily guaranteed that 
Eq.~(\ref{deln}) is valid in such cases.
Moreover the estimation (\ref{deln}) does not 
take into account the evolution of $\delta_m$ after $z=z_\Lambda$
to the present epoch ($z=0$).
In order to see the validity of the formula (\ref{deln}) we have evaluated
numerical values of $\delta n (t_\Lambda)$ as well as
$\delta n (t_0)$ integrated up to the present epoch.
{}From Table II we find that the estimation (\ref{deln}) agrees well 
with the numerical obtained $\delta n (t_\Lambda)$ for $p \le 5$.
The difference appears for $p \ge 6$, but it is not significant.

After the system enters the epoch of an accelerated 
expansion,
the momentum $k$ can again become smaller than $aM$.
Hence the $k$-dependence is not necessarily negligible
even for $z<z_\Lambda$.
However we find from Table II that 
$\delta n (t_0)$ is not much different from $\delta n (t_\Lambda)$
derived by Eq.~(\ref{deln}).
Thus the analytic estimation (\ref{deln}) is certainly reliable 
to place constraints on model parameters except for $p \gg 1$.

Observationally we do not find any strong signature
for the difference of slopes of the spectra of LSS 
and CMB \cite{Teg}.
If we take the mild bound $\delta n \lesssim 0.05$, we obtain 
the constraint $p \ge 5$. It is interesting to recall that 
LGC are well satisfied for $p \ge 5$.

\begin{figure}
\begin{centering}\includegraphics[width=3.4in,height=3.2in]{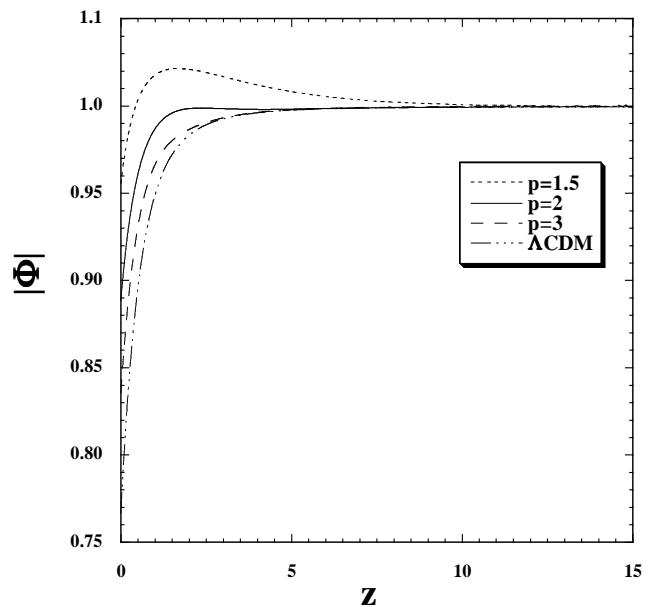} 
\par\end{centering}
\caption{\label{fig7} The evolution of the effective gravitational potential 
$\Phi \equiv (\alpha-\varphi)/2$
in the model $m(r)=(-r-1)^p$ for the mode $k/a_0H_0=4$
with three different values of $p$. 
Note that $\Phi$ is initially normalized as unity.
We also show the evolution of $\Phi$ in the $\Lambda$CDM model.
For larger $p$ the decay of the gravitational potential is more significant.
When $p \ge 2$ the models are well consistent with CMB low-multipole data. }
\end{figure}

Finally we shall discuss the integrated Sachs-Wolfe (ISW) effect in the model 
$m(r)=C(-r-1)^p$. In order to confront the model with CMB it is convenient to 
study the evolution of an effective gravitational potential $\Phi \equiv 
(\alpha-\varphi)/2$ in the longitudinal gauge 
($\chi=0$) \cite{Song,linear}. 
Under the sub-horizon approximation used in 
Refs.~\cite{efp,Tsujiper} we obtain, from 
Eqs.~(\ref{pereq5}) and (\ref{pereq6}), the following relation 
\begin{eqnarray}
\Phi \simeq -\frac32 \frac{a^2H^2}{k^2}
\Omega_m \delta_m\,.
\end{eqnarray}
In the $\Lambda$CDM model the gravitational potential 
remains constant during the standard matter era, but it
decays after the system enters the accelerated epoch.
This leads to the ISW effect for 
low multipoles of  the CMB power spectrum.
In our $f(R)$ model (\ref{ourmodel}), the additional growth of 
matter perturbations in the region $z<z_k$ 
changes the evolution of $\Phi$.

In Fig.~\ref{fig7} we plot the evolution of $\Phi$ in the models 
$m(r)=(-r-1)^p$ with the mode $k/a_0H_0=4$ 
for several different values of $p$.
For smaller $p$ the gravitational potential does not decay much 
because the duration of the region $z<z_k$ gets longer.
The models that cancel the ISW effect by the additional growth 
of $\delta_m$ are consistent with the CMB data \cite{SPH},
which means that the models with $p \ge 2$ are allowed.
Hence the information coming from the ISW effect of 
CMB does not provide strong constraints on the model 
parameters compared to the galaxy spectrum 
discussed above.

%%%%%%%%%%%%%%
\section{Conclusions}
\label{conclude}
%%%%%%%%%%%%%%

We have discussed various observational signatures of $f(R)$
dark energy scenarios that satisfy both cosmological and local 
gravity constraints.
The $f(R)$ models do not have much freedom to fulfill
all such constraints. Generally the models need to mimic the 
$\Lambda$CDM model in a high-curvature region where
local gravity experiments are carried out ($R \gg R_0 \sim H_0^2$).
The deviation from the $\Lambda$CDM cosmology becomes
important after the end of the matter-dominated epoch 
in such viable models.

The models given in Eqs.~(\ref{mo1}) and (\ref{mo2}) belong to 
such viable classes that possess a de-Sitter attractor at $R=R_1>0$ 
and satisfy the condition $f(R=0)=0$. 
The stability condition requires that $f_{,R}>0$ and $f_{,RR}>0$
for $R \ge R_1$. In the models (\ref{mo1}) and (\ref{mo2})
either of these conditions is violated in the region $R<R_1$, 
but this is not problematic as long as 
the Ricci scalar does not oscillate. 
In Sec.~\ref{moex} we discussed general properties
about viable $f(R)$ models and proposed 
another simple model (\ref{mo3}) that can satisfy 
the conditions $f_{,R}>0$ and $f_{,RR}>0$ 
for all positive $R$.

The model we have studied in this paper is given by 
$m(r)=C(-r-1)^p~(C>0, p>1)$,
where $m$ and $r$ are defined by Eq.~(\ref{mdef}).
In the high-curvature region characterized by $R \gg R_0$, 
these recover the models (\ref{mo1}), (\ref{mo2}) with finite $p$
and also the model (\ref{mo3}) with $p \to \infty$.
Moreover the departure from the $\Lambda$CDM 
cosmology can be captured by the growth of the quantity $m$
as the solutions approach the de-Sitter fixed point 
on the line $r=-2$. 
In fact this model possesses rich observational signatures 
relevant to SN Ia, galaxy clustering and CMB.

The equation of state of dark energy shows a peculiar 
divergent behavior with a cosmological boundary crossing.
When the deviation from the $\Lambda$CDM model is 
significant, the redshift $z_c$ at which such a divergence 
occurs can be as close as a few. 
This may be detectable in future 
observations of SN Ia and weak lensing.
Using observational bounds on the 
equation of state of dark energy at present ($z=0$),
we found that the models with $p \ge 2$ are consistent 
with the data. The deviation parameter $m$ is constrained
to be $m(z=0) \lesssim 0.3$.
We also showed that the models with $p \ge 2$ satisfy 
sound horizon constraints of CMB. 

We have discussed the evolution of matter density perturbations
$\delta_m$ together with the perturbation in the Ricci scalar $R$.
The mass squared of the scalaron, $M^2 \simeq 1/(3f_{,RR})$, is much 
larger than $H^2$ and can cross the value $k^2/a^2$
during the matter era. 
In the early epoch with $M^2 \gg k^2/a^2$ the matter perturbation
evolves as in the standard way, provided that the oscillating mode
$\delta R_{\rm osc}$  (scalaron) is suppressed relative to 
the induced matter mode $\delta R_{\rm ind}$. 
However the scalaron dominates at early 
epochs unless the coefficient of this mode is chosen to be very small 
so that $\delta R$ is always as close as $\delta R_{\rm ind}$. 
The dominance of the scalaron means the violation of the condition
(\ref{con1}), which leads to an instability of solutions 
in the matter-dominated epoch.
This property persists in the radiation era and hence it poses a serious
problem about how an over-production of the scalaron is avoided
in the early universe.

In the late epoch characterized by $M^2 \ll k^2/a^2$ the matter perturbation evolves 
in a non-standard way, see Eq.~(\ref{delm2}). This leads to additional 
growth of matter perturbations depending on the wavenumber $k$.
Considering the evolution of $\delta_m$ by the time $t_\Lambda$
at which an  accelerated expansion sets in, 
the difference about spectral indices of the power spectra between galaxy 
clustering and CMB is given by Eq.~(\ref{deln}). 
We also integrated perturbation equations numerically by the present epoch
and found that the estimation (\ref{deln}) agrees fairly well with 
numerical values.
Using the rather mild criterion $\delta n \lesssim 0.05$, the constraint on the parameter
$p$ is given by $p \ge 5$. 
We have also studied the evolution of an effective gravitational potential 
and found that the integrated Sachs-Wolfe effect in low multipoles of CMB
does not provide a stronger constraint than the one coming from LSS.

It will be certainly of interest to place stringent constraints on the model
parameters using future high-precision observational data. We hope that this 
allows us to find some signatures about the deviation from 
the $\Lambda$CDM model. In particular the detection of unusual 
behavior of the equation of state of dark energy can be strong 
evidence for $f(R)$ gravity models.

%%%%%%%%%%%%%%%%%%%%%%%%%%%%%%%%%%%%%
\section*{ACKNOWLEDGEMENTS}
I thank Luca Amendola,  Nicola Bartolo, Daniel Bertacca, 
Antonio De Felice, Martin Kunz, Sabino Matarrese, David Polarski, 
Alexei Starobinsky, Reza Tavakol and Kotub Uddin 
for very useful discussions. 
I am also grateful to Roy Maartens, Reza Tavakol,  
Luca Amendola and Sabino Matarrese
for supporting visits to University of Portsmouth, University of Queen Mary,
Rome observatory and University of Padova 
during which this work was completed.
%%%%%%%%%%%%%%%%%%%%%%%%%%%%%%%%%%%%%

%
\section*{{\large Appendix}}

In this Appendix we present the perturbation 
equations convenient for numerical purpose.
We neglect the contribution of radiation, i.e., $x_4=0$.
Using the dimensionless variables given in Eq.~(\ref{didef}) and
introducing a new quantity $\delta \tilde{F}=\delta F/F$, 
Eqs.~(\ref{delm}) and (\ref{delF}) can be written as 
\begin{eqnarray}
\label{per1nu}
& &\delta_m''+\left(x_3-\frac12 x_1 \right)\delta_m'-
\frac32 (1-x_1-x_2-x_3) \delta_m \nonumber \\
& &=\frac12 \biggl[ \left\{ \frac{k^2}{x_5^2}-6
+3x_1^2-3x_1'-3x_1 (x_3-1) \right\}
\delta \tilde{F}\nonumber \\ 
& &~~~~~~~+3(-2x_1+x_3-1)\delta \tilde{F}'+
3\delta \tilde{F}'' \biggr]\,,\\
\label{per2nu}
& & \delta \tilde{F}''+(1-2x_1+x_3)\delta \tilde{F}' 
\nonumber \\
& & +\left[ \frac{k^2}{x_5^2}-2x_3+\frac{2x_3}{m}
-x_1 (x_3+1)-x_1'+x_1^2 \right]\delta \tilde{F} 
\nonumber \\
& & =(1-x_1-x_2-x_3)\delta _m-x_1\delta_m'\,,
\end{eqnarray}
where a new variable, $x_5 \equiv aH$, satisfies
\begin{eqnarray}
\label{x5eq}
x_5'=(x_3-1)x_5\,.
\end{eqnarray}
Note that the perturbation $\delta R$ is given by
\begin{eqnarray}
\delta R=6H^2 \frac{x_3}{m} \delta \tilde{F}\,.
\end{eqnarray}
The evolution of the Hubble parameter is known by solving 
the equation 
\begin{eqnarray}
\label{Hdeq}
\frac{H'}{H}=x_3-2\,.
\end{eqnarray}
Solving equations (\ref{per1nu}) and (\ref{per2nu}) together with the 
background equations (\ref{N1})-(\ref{N3}), (\ref{x5eq}) 
and (\ref{Hdeq}) numerically, 
we find the evolution of $\delta_m$ and $\delta R$.

%%%%%%%%%%%%%%%%%%


\begin{thebibliography}{10}
%%%%%%%%%%%%%%%%%%

\bibitem{review} 
V.~Sahni and A.~A.~Starobinsky, 
%``The Case for a Positive Cosmological Lambda-term,''
Int.\ J.\ Mod.\ Phys.\ D \textbf{9}, 373 (2000); 
S.~M.~Carroll,
%``The cosmological constant,''
Living Rev.\ Rel.\  \textbf{4}, 1 (2001); V.~Sahni, %``Dark matter and dark energy,''
Lect.\ Notes Phys.\  \textbf{653}, 141 (2004) {[}arXiv:astro-ph/0403324];
T.~Padmanabhan, %``Cosmological constant: The weight of the vacuum,''
Phys.\ Rept.\  \textbf{380}, 235 (2003); 
P.~J.~E.~Peebles and B.~Ratra, 
%``The cosmological constant and dark energy,''
Rev.\ Mod.\ Phys.\  \textbf{75}, 559 (2003); 
S.~Nojiri and S.~D.~Odintsov, Int.\ J.\ Geom.\ Meth.\ Mod.\ Phys.\ 
\textbf{4}, 115 (2007).

\bibitem{CST} 
E.~J.~Copeland, M.~Sami and S.~Tsujikawa, 
%``Dynamics of dark energy,''
Int.\ J.\ Mod.\ Phys.\ D \textbf{15}, 1753 (2006).

\bibitem{star} 
A.~A.~Starobinsky, 
%``A new type of isotropic cosmological models without singularity,''
Phys.\ Lett.\ B \textbf{91}, 99 (1980).

\bibitem{fR} 
S.~Capozziello, V.~F.~Cardone, S.~Carloni and A.~Troisi,
Int.\ J.\ Mod.\ Phys.\ D \textbf{12}, 1969 (2003); 
S.~Capozziello, S.~Carloni and A.~Troisi,
%``Quintessence without scalar fields,''
arXiv:astro-ph/0303041;
S.~M.~Carroll,
V.~Duvvuri, M.~Trodden and M.~S.~Turner, 
Phys.\ Rev.\ D 70,
043528 (2004).

\bibitem{Capoluca} 
S.~Capozziello, F.~Occhionero and L.~Amendola,
Int.\ J.\ Mod.\ Phys.\ D \textbf{1} (1993) 615.

\bibitem{early} 
S.~Nojiri and S.~D.~Odintsov, Phys.\ Rev.\ D
\textbf{68}, 123512 (2003); 
M.~E.~Soussa and R.~P.~Woodard,
Gen.\ Rel.\ Grav.\  \textbf{36}, 855 (2004); 
G.~Allemandi, A.~Borowiec and M.~Francaviglia, 
%``Accelerated cosmological models in Ricci squared gravity,''
Phys.\ Rev.\ D \textbf{70}, 103503 (2004); 
D.~A.~Easson, 
%``Cosmic Acceleration and Modified Gravitational Models,''
Int.\ J.\ Mod.\ Phys.\ A \textbf{19}, 5343 (2004); 
S.~M.~Carroll \textit{et al.}, 
%``The cosmology of generalized modified gravity models,''
Phys.\ Rev.\ D \textbf{71}, 063513 (2005); 
S.~Carloni, P.~K.~S.~Dunsby, S.~Capozziello and A.~Troisi, 
%``Cosmological dynamics of R**n gravity,''
Class.\ Quant.\ Grav.\  \textbf{22}, 4839 (2005).

\bibitem{Dolgov} 
A.~D.~Dolgov and M.~Kawasaki, Phys.\ Lett.\ B \textbf{573}, 1 (2003).

\bibitem{LG} 
T.~Chiba, 
%``1/R gravity and scalar-tensor gravity,''
Phys.\ Lett.\ B \textbf{575}, 1 (2003).

\bibitem{APT} 
L.~Amendola, D.~Polarski and S.~Tsujikawa, 
%``Are $f(R)$ dark energy models cosmologically viable?,''
Phys.\ Rev.\ Lett.\  \textbf{98}, 131302 (2007);
arXiv:astro-ph/0605384.

\bibitem{recent} 
S.~Capozziello, S.~Nojiri, S.~D.~Odintsov and A.~Troisi, 
Phys.\ Lett.\  B \textbf{639}, 135 (2006); 
A.~W.~Brookfield, C.~van de Bruck and L.~M.~H.~Hall, 
%``Viability of f(R) theories with additional powers of curvature,''
Phys.\ Rev.\ D \textbf{74}, 064028 (2006); 
M.~Amarzguioui, O.~Elgaroy, D.~F.~Mota and T.~Multamaki,
%``Cosmological constraints on $f(R)$ gravity theories 
%within the Palatini approach,''
Astron.\ Astrophys.\  {\bf 454}, 707 (2006);
T.~P.~Sotiriou,
%``$f(R)$ gravity and scalar-tensor theory,''
Class.\ Quant.\ Grav.\  {\bf 23}, 5117 (2006);
S.~Nojiri and S.~D.~Odintsov,
Phys.\ Rev.\  D {\bf 74}, 086005 (2006);
N.~J.~Poplawski,
%``Interacting dark energy in $f(R)$ gravity,''
Phys.\ Rev.\  D {\bf 74}, 084032 (2006);
A.~Borowiec, W.~Godlowski and M.~Szydlowski, 
%``Accelerated cosmological models in modified gravity tested by distant
%supernovae SNIa data,''
Phys.\ Rev.\  D \textbf{74}, 043502 (2006);
T.~Koivisto,
%``The matter power spectrum in $f(R)$ gravity,''
Phys.\ Rev.\  D {\bf 73}, 083517 (2006);
A.~de la Cruz-Dombriz and A.~Dobado, 
%``A $f(R)$ gravity without cosmological constant,''
Phys.\ Rev.\  D \textbf{74}, 087501 (2006);  
T.~Multamaki and I.~Vilja,
Phys.\ Rev.\  D {\bf 74}, 064022 (2006);
T.~P.~Sotiriou, 
%``Curvature scalar instability in f(R) gravity,''
Phys.\ Lett.\  B \textbf{645}, 389 (2007); 
T.~P.~Sotiriou and S.~Liberati,
%``Metric-affine $f(R)$ theories of gravity,''
Annals Phys.\  \textbf{322}, 935 (2007); 
V.~Faraoni and S.~Nadeau,
%``The (pseudo)issue of the conformal frame revisited,''
Phys.\ Rev.\  D \textbf{75}, 023501 (2007); 
D.~Huterer and E.~V.~Linder,
Phys.\ Rev.\  D \textbf{75}, 023519 (2007); 
S.~Fay, R.~Tavakol and S.~Tsujikawa, 
Phys.\ Rev.\  D \textbf{75}, 063509 (2007);
K.~Kainulainen, J.~Piilonen, V.~Reijonen and D.~Sunhede,
%``Spherically symmetric spacetimes in f(R) gravity theories,''
Phys.\ Rev.\  D {\bf 76}, 024020 (2007);
A.~De Felice and M.~Hindmarsh,
%``Unsuccessful cosmology with Modified Gravity Models,''
JCAP {\bf 0706}, 028 (2007);
K.~Uddin, J.~E.~Lidsey and R.~Tavakol,
%``Cosmological perturbations in Palatini modified gravity,''
Class.\ Quant.\ Grav.\  {\bf 24}, 3951 (2007);
J.~C.~C.~de Souza and V.~Faraoni,
%``The phase space view of f(R) gravity,''
Class.\ Quant.\ Grav.\  {\bf 24}, 3637 (2007);
M.~S.~Movahed, S.~Baghram and S.~Rahvar,
Phys.\ Rev.\  D {\bf 76}, 044008 (2007);
E.~O.~Kahya and V.~K.~Onemli, 
%``Quantum stability of a w < -1 phase of cosmic acceleration,''
arXiv:gr-qc/0612026; 
M.~Fairbairn and S.~Rydbeck, 
%``Expansion history and f(R) modified gravity,''
arXiv:astro-ph/0701900; 
P.~J.~Zhang, 
Phys.\ Rev.\  D \textbf{73}, 123504 (2006);
arXiv:astro-ph/0701662;  
G.~Cognola, M.~Gastaldi and S.~Zerbini, 
arXiv:gr-qc/0701138; 
D.~Bazeia, B.~Carneiro da Cunha, R.~Menezes and A.~Y.~Petrov, 
arXiv:hep-th/0701106; 
T.~Rador, arXiv:hep-th/0701267; 
S.~Bludman, arXiv:astro-ph/0702085;
L.~M.~Sokolowski, arXiv:gr-qc/0702097; 
S.~Fay, S.~Nesseris and L.~Perivolaropoulos, 
%``Can f(R) Modified Gravity Theories Mimic a LCDM Cosmology?,''
arXiv:gr-qc/0703006; 
S.~Nojiri, S.~D.~Odintsov and P.~V.~Tretyakov,
%``Dark energy from modified F(R)-scalar-Gauss-Bonnet gravity,''
arXiv:0704.2520 {[}hep-th]; 
O.~Bertolami, C.~G.~Boehmer, T.~Harko and F.~S.~N.~Lobo, 
%``Extra force in f(R) modified theories of gravity,''
arXiv:0704.1733 {[}gr-qc];
S.~Capozziello and M.~Francaviglia,
%``Extended Theories of Gravity and their Cosmological 
%and Astrophysical Applications,''
arXiv:0706.1146 [astro-ph];
A.~De Felice, P.~Mukherjee and Y.~Wang,
%``Observational Bounds on Modified Gravity Models,''
arXiv:0706.1197 [astro-ph];
B.~Li, J.~D.~Barrow and D.~F.~Mota,
arXiv:0707.2664 [gr-qc];
K.~Bamba, Z.~K.~Guo and N.~Ohta,
arXiv:0707.4334 [hep-th];
C.~G.~Boehmer, T.~Harko and F.~Lobo,
arXiv:0709.0046 [gr-qc];
C.~G.~Boehmer, T.~Harko and F.~S.~N.~Lobo,
%``Dark matter as a geometric effect in f(R) gravity,''
arXiv:0709.0046 [gr-qc].

\bibitem{AGPT} 
L.~Amendola, R.~Gannouji, D.~Polarski and S.~Tsujikawa,
%``Conditions for the cosmological viability of f(R) 
%dark energy models,''
Phys.\ Rev.\ D \textbf{75}, 083504 (2007).

\bibitem{Li} 
B.~Li and J.~D.~Barrow,
%``The Cosmology of f(R) Gravity in Metric Variational Approach,''
Phys.\ Rev.\  D {\bf 75}, 084010 (2007).

\bibitem{AT07}
L.~Amendola and S.~Tsujikawa,
%``Phantom crossing, equation-of-state singularities, 
%and local gravity constraints in $f(R)$ models,''
arXiv:0705.0396 [astro-ph].

\bibitem{Hu}
W.~Hu and I.~Sawicki,
%``Models of f(R) Cosmic Acceleration that Evade Solar-System Tests,''
Phys.\ Rev.\  D {\bf 76}, 064004 (2007).

\bibitem{star07}
A.~A.~Starobinsky,
%``Disappearing cosmological constant in f(R) gravity,''
JETP Lett.\  {\bf 86}, 157 (2007).

\bibitem{Appleby}
S.~A.~Appleby and R.~A.~Battye,
%``Do consistent $F(R)$ models mimic General 
%Relativity plus $\Lambda$?,''
Phys.\ Lett.\  B {\bf 654}, 7 (2007).

\bibitem{KW}
J.~Khoury and A.~Weltman,
%``Chameleon fields: Awaiting surprises for tests of gravity in space,''
Phys.\ Rev.\ Lett.\  {\bf 93}, 171104 (2004);
Phys.\ Rev.\  D {\bf 69}, 044026 (2004).

\bibitem{Olmo}
G.~J.~Olmo,
Phys.\ Rev.\  D {\bf 72}, 083505 (2005).

\bibitem{lgcpapers}
G.~J.~Olmo,
%``The gravity lagrangian according to solar system experiments,''
Phys.\ Rev.\ Lett.\  {\bf 95}, 261102 (2005);
A.~L.~Erickcek, T.~L.~Smith and M.~Kamionkowski,
%``Solar system tests do rule out 1/R gravity,''
Phys.\ Rev.\  D {\bf 74}, 121501 (2006);
V.~Faraoni,
%``Solar system experiments do not yet veto modified gravity models,''
Phys.\ Rev.\  D {\bf 74}, 023529 (2006);
A.~F.~Zakharov, A.~A.~Nucita, F.~De Paolis and G.~Ingrosso, 
%``Solar system constraints on R$^n$ gravity,''
Phys.\ Rev.\  D \textbf{74}, 107101 (2006); 
I.~Navarro and K.~Van Acoleyen,
%``f(R) actions, cosmic acceleration and local tests of gravity,''
JCAP {\bf 0702}, 022 (2007);
T.~Chiba, T.~L.~Smith and A.~L.~Erickcek,
%``Solar System constraints to general f(R) gravity,''
Phys.\ Rev.\  D {\bf 75}, 124014 (2007);
G.~Allemandi and M.~L.~Ruggiero,
%``Constraining Extended Theories of Gravity using Solar System Tests,''
arXiv:astro-ph/0610661;
X.~H.~Jin, D.~J.~Liu and X.~Z.~Li,
%``Solar System tests disfavor $f(R)$ gravities,''
arXiv:astro-ph/0610854;
T.~Faulkner, M.~Tegmark, E.~F.~Bunn and Y.~Mao,
arXiv:astro-ph/0612569; 
S.~Nojiri and S.~D.~Odintsov,
arXiv:0707.1941 [hep-th].

\bibitem{Fara}
V.~Faraoni, 
%``Matter instability in modified gravity,''
Phys.\ Rev.\  D \textbf{74}, 104017 (2006).

\bibitem{Song} 
Y.~S.~Song, W.~Hu and I.~Sawicki,
%``The large scale structure of f(R) gravity,''
Phys.\ Rev.\  D {\bf 75}, 044004 (2007);
I.~Sawicki and W.~Hu,
%``Stability of Cosmological Solution in f(R) Models of Gravity,''
Phys.\ Rev.\  D {\bf 75}, 127502 (2007).

\bibitem{rbean} 
N.~Agarwal and R.~Bean,
%``The dynamical viability of scalar-tensor theories 
%with a general coupling,''
arXiv:0708.3967 [astro-ph].

\bibitem{hoyle}
C.~D.~Hoyle, D.~J.~Kapner, B.~R.~Heckel, E.~G.~Adelberger,
J.~H.~Gundlach, U.~Schmidt and H.~E.~Swanson, 
%``Sub-millimeter tests of the gravitational inverse-square law,''
Phys.\ Rev.\ D \textbf{70}, 042004 (2004).

\bibitem{Shapiro}
I.~I.~Shapiro,
%``Fourth Test of General Relativity,''
Phys.\ Rev.\ Lett.\  {\bf 13}, 789 (1964).

\bibitem{Nese} 
S.~Nesseris and L.~Perivolaropoulos, 
%``Crossing the phantom divide: Theoretical implications 
%and observational status,''
JCAP \textbf{0701}, 018 (2007); 
Phys.\ Rev.\  D \textbf{75}, 023517 (2007).

\bibitem{Astier} 
P.~Astier \textit{et al.}, 
%``The Supernova Legacy Survey: Measurement of Omega_M, 
%Omega_Lambda and w from the First Year Data Set,''
Astron.\ Astrophys.\  \textbf{447}, 31 (2006).

\bibitem{Spergel} 
D.~N.~Spergel \textit{et al.}, 
%``Wilkinson Microwave Anisotropy Probe (WMAP) three year results:
%Implications for cosmology,''
Astrophys.\ J.\ Suppl.\  {\bf 170}, 377 (2007).

\bibitem{coupled}
L.~Amendola, Phys. Rev. {\bf D62}, 043511 (2000).

\bibitem{Arman}
A.~Shafieloo, U.~Alam, V.~Sahni and A.~A.~Starobinsky,
%``Smoothing supernova data to reconstruct the 
%expansion history of the universe,''
Mon.\ Not.\ Roy.\ Astron.\ Soc.\  {\bf 366}, 1081 (2006);
A.~Shafieloo, arXiv:astro-ph/0703034.

\bibitem{metric}
H.~Kodama and M.~Sasaki,
%``Cosmological Perturbation Theory,''
Prog.\ Theor.\ Phys.\ Suppl.\  {\bf 78}, 1 (1984);
V.~F.~Mukhanov, H.~A.~Feldman and R.~H.~Brandenberger,
Phys.\ Rept.\  {\bf 215}, 203 (1992);
B.~A.~Bassett, S.~Tsujikawa and D.~Wands,
%``Inflation dynamics and reheating,''
Rev.\ Mod.\ Phys.\  {\bf 78}, 537 (2006).

\bibitem{Hwang}
J.~c.~Hwang and H.~Noh,
Phys.\ Rev.\  D {\bf 71}, 063536 (2005);
Phys.\ Rev.\  D {\bf 66}, 084009 (2002).	

\bibitem{linear} 
S.~M.~Carroll, I.~Sawicki, A.~Silvestri and M.~Trodden, 
%``Modified-Source Gravity and Cosmological Structure Formation,''
New J.\ Phys.\  \textbf{8}, 323 (2006); 
R.~Bean, D.~Bernat, L.~Pogosian, A.~Silvestri and M.~Trodden, 
%``Dynamics of Linear Perturbations in f(R) Gravity,''
Phys.\ Rev.\  D \textbf{75}, 064020 (2007);
L.~Pogosian and A.~Silvestri,
%``The pattern of growth in viable f(R) cosmologies,''
arXiv:0709.0296 [astro-ph];
I.~Laszlo and R.~Bean,
%``Non-linear growth in modified gravity,''
arXiv:0709.0307 [astro-ph].

\bibitem{efp} 
B.~Boisseau, G.~Esposito-Farese, D.~Polarski and A.~A.~Starobinsky, 
%``Reconstruction of a scalar-tensor theory of gravity in 
%an accelerating universe,''
Phys.\ Rev.\ Lett.\  \textbf{85}, 2236 (2000); 
G.~Esposito-Farese and D.~Polarski, 
%``Scalar-tensor gravity in an accelerating universe,''
Phys.\ Rev.\  D \textbf{63}, 063504 (2001).

\bibitem{Tsujiper} 
S.~Tsujikawa,
%``Matter density perturbations and effective gravitational constant in
%modified gravity models of dark energy,''
Phys.\ Rev.\  D {\bf 76}, 023514 (2007).

\bibitem{SPH} 
Y.~S.~Song, H.~Peiris and W.~Hu,
%``Cosmological Constraints on f(R) Acceleration Models,''
Phys.\ Rev.\  D {\bf 76}, 063517 (2007).

\bibitem{Teg} 
M.~Tegmark {\it et al.},
%``Cosmological Constraints from the SDSS Luminous Red Galaxies,''
Phys.\ Rev.\  D {\bf 74}, 123507 (2006).

\end{thebibliography}
\end{document}